\documentclass[11pt]{article}
\usepackage{moriond,epsfig,graphicx}

\makeatletter
\let\jnl@style=\rmfamily 
\def\ref@jnl#1{{\jnl@style#1}}%
\newcommand\aj{\ref@jnl{AJ}}%
\newcommand\araa{\ref@jnl{ARA\&A}}%
\newcommand\apj{\ref@jnl{ApJ}}%
\newcommand\apjl{\ref@jnl{ApJ}}%
\newcommand\apjs{\ref@jnl{ApJS}}%
\newcommand\ao{\ref@jnl{Appl.~Opt.}}%
\newcommand\apss{\ref@jnl{Ap\&SS}}%
\newcommand\aap{\ref@jnl{A\&A}}%
\newcommand\aapr{\ref@jnl{A\&A~Rev.}}%
\newcommand\aaps{\ref@jnl{A\&AS}}%
\newcommand\azh{\ref@jnl{AZh}}%
\newcommand\baas{\ref@jnl{BAAS}}%
\newcommand\jrasc{\ref@jnl{JRASC}}%
\newcommand\memras{\ref@jnl{MmRAS}}%
\newcommand\mnras{\ref@jnl{MNRAS}}%
\newcommand\pra{\ref@jnl{Phys.~Rev.~A}}%
\newcommand\prb{\ref@jnl{Phys.~Rev.~B}}%
\newcommand\prc{\ref@jnl{Phys.~Rev.~C}}%
\newcommand\prd{\ref@jnl{Phys.~Rev.~D}}%
\newcommand\pre{\ref@jnl{Phys.~Rev.~E}}%
\newcommand\prl{\ref@jnl{Phys.~Rev.~Lett.}}%
\newcommand\pasp{\ref@jnl{PASP}}%
\newcommand\pasj{\ref@jnl{PASJ}}%
\newcommand\qjras{\ref@jnl{QJRAS}}%
\newcommand\skytel{\ref@jnl{S\&T}}%
\newcommand\solphys{\ref@jnl{Sol.~Phys.}}%
\newcommand\sovast{\ref@jnl{Soviet~Ast.}}%
\newcommand\ssr{\ref@jnl{Space~Sci.~Rev.}}%
\newcommand\zap{\ref@jnl{ZAp}}%
\newcommand\nat{\ref@jnl{Nature}}%
\newcommand\iaucirc{\ref@jnl{IAU~Circ.}}%
\newcommand\aplett{\ref@jnl{Astrophys.~Lett.}}%
\newcommand\apspr{\ref@jnl{Astrophys.~Space~Phys.~Res.}}%
\newcommand\bain{\ref@jnl{Bull.~Astron.~Inst.~Netherlands}}%
\newcommand\fcp{\ref@jnl{Fund.~Cosmic~Phys.}}%
\newcommand\gca{\ref@jnl{Geochim.~Cosmochim.~Acta}}%
\newcommand\grl{\ref@jnl{Geophys.~Res.~Lett.}}%
\newcommand\jcp{\ref@jnl{J.~Chem.~Phys.}}%
\newcommand\jgr{\ref@jnl{J.~Geophys.~Res.}}%
\newcommand\jqsrt{\ref@jnl{J.~Quant.~Spec.~Radiat.~Transf.}}%
\newcommand\memsai{\ref@jnl{Mem.~Soc.~Astron.~Italiana}}%
\newcommand\nphysa{\ref@jnl{Nucl.~Phys.~A}}%
\newcommand\physrep{\ref@jnl{Phys.~Rep.}}%
\newcommand\physscr{\ref@jnl{Phys.~Scr}}%
\newcommand\planss{\ref@jnl{Planet.~Space~Sci.}}%
\newcommand\procspie{\ref@jnl{Proc.~SPIE}}%

\makeatother

\def\l5{CL5}
\def\cl6{CL6}
\def\cl7{CL7}
\def\cl9{CL9}
\def\cl10{CL10}
\def\cl11{CL11}
\def\cl14{CL14}
\def\cl24{CL24}
\def\l101{CL101}
\def\cl102{CL102}
\def\cl103{CL103}
\def\cl104{CL104}
\def\cl105{CL105}
\def\cl106{CL106}
\def\cl107{CL107}
\def\cl108{CL108}

\def\Lx{L_{\rm x}}
\def\Yx{Y_{\mkern-0.2mu\rm x}}
\def\Tx{T_{\mkern-0.4mu\rm x}}
\def\Mg{M_{\rm g}}
\def\Mtot{M_{\rm tot}}
\def\M500{M_{\rm 500}}

\def\hide#1{}

\begin{document}

\title{``The Perfect Slope'': A new robust low-scatter X-ray mass
  indicator\\
  for clusters of galaxies}

\author{Alexey Vikhlinin, Andrey V. Kravtsov, Daisuke Nagai}

\address{Harvard-Smithsonian Center for Astrophysics\\
  The University of Chicago\\
  Caltech}

\maketitle

\abstracts{This presentation is a Moriond version of our recent paper
  (Kravtsov, Vikhlinin \& Nagai~\cite{2006astro.ph..3205K}) where we
  discussed X-ray proxies for the total cluster mass, including the
  spectral temperature ($\Tx$), gas mass measured within $r_{500}$
  ($\Mg$), and the new proxy, $\Yx$, which is a simple product of $\Tx$
  and $\Mg$. We use mock \emph{Chandra} images constructed for a sample
  of clusters simulated with high resolution in the concordance
  $\Lambda$CDM cosmology. The simulated clusters exhibit tight
  correlations between the considered observables and total mass. The
  normalizations of the $M_{500}-\Tx$, $\Mg-\Tx$, and $M_{500}-\Yx$
  relations agree to better than $\approx 10-15\%$ with the current
  observational measurements of these relations. Our results show that
  $\Yx$ is the best mass proxy with a remarkably low scatter of only
  $\approx 5-7\%$ in $\M500$ for a fixed $\Yx$, at both low and high
  redshifts and regardless of whether clusters are relaxed or not.  In
  addition, we show that redshift evolution of the $\Yx-M_{500}$
  relation is close to the self-similar prediction, which makes $\Yx$ a
  very attractive mass indicator for measurements of the cluster mass
  function from X-ray selected samples.  }

\section{Introduction}
\label{sec:intro}

The evolution of the cluster abundance is one of the most sensitive
probes of cosmology. The potential and importance of this method have
motivated efforts to construct several large surveys of high-redshift
clusters during the next several years. However, in order to realize the
full statistical power of the upcoming cluster surveys, it is paramount
that the relation between cluster mass and observables and any potential
biases are well known. 

Several cluster observables based on the galaxy velocities, optical
light, X-ray observables such as luminosity, temperature, mass of the
intracluster medium (ICM), and Sunyaev-Zel'dovich flux have been
proposed as proxies of the total cluster mass (see a recent review by
Voit~\cite{voit05}). In this study we focus on the mass indicators
derived from cluster X-ray observables. 

X-ray luminosity is the most straightforward mass indicator to measure
observationally. However, $\Lx$ is also the least accurate (internally)
of all proposed X-ray proxies for $\Mtot$ with a large
scatter~\cite{david_etal93,stanek_etal06} and deviations of the slope of
the $\Lx-M$ relation from the self-similar
prediction~\cite{allen_etal03}.  The most common choice of mass proxy in
the cluster cosmological studies has been the X-ray temperature of the
intracluster gas
\cite{henry_arnaud91,oukbir_blanchard92,markevitch98,ikebe_etal02}. The
scatter in the $M-\Tx$ relation is smaller compared to that in the
$\Lx-M$ relation (the upper limit from observations is $\approx15\%$ in
$M$ for fixed $T$ for relaxed clusters~\cite{vikhlinin_etal06}). In
general, existence of a tight relation such as $M-\Tx$ indicates that
clusters are regular population of objects with their global properties
tightly related to total mass, and scatter caused by secondary effects
such as substructure in the ICM, non-gravitational processes, and
mergers \cite{ohara_etal06}. More recently, gas mass was used as a proxy
for $\Mtot$ \cite{vikhlinin_etal03,voevodkin_vikhlinin04}. The practical
advantage of $\Mg$ over $\Tx$ is that it can be measured from the X-ray
imaging alone. Also, $M_g$ can be expected to be less sensitive to
mergers which should translate into smaller scatter in the $M_{\rm g}-M$
relation. The caveat is that trend of gas mass with cluster mass and
evolution with redshift are not yet fully understood. 

The use of clusters as efficient probes for precision cosmology puts
stringent requirements on observable cluster mass proxies: 1) tight,
low-scatter correlation between the proxy and mass, with the scatter
insensitive to mergers etc.{}, and 2) simple power-law relation and
evolution which can be described by a small number of parameters and be
as close as possible to the prediction of the self-similar model. The
last point is crucial to ensure that the self-calibration strategies for
analyses of large cluster surveys
\cite{levine_etal02,hu03,majumdar_mohr03,lima_hu05,wang_etal04} are
successful. This is because self-calibration is powerful when cluster
scaling relations and their evolution have a simple form which can be
parameterized with a small number of parameters. 

In general, a mass proxy does not have to be a single cluster property,
such as $\Lx$, $\Tx$ or $\Mg$. Any physically-motivated combination of
these variables that is expected to be tightly related to cluster mass
can be used to construct a valid mass indicator. A hint for a better
X-ray mass proxy is provided by recent studies based on cosmological
simulations of cluster formation \cite{motl_etal05,nagai06}, which show
that integrated SZ flux, $Y_{\rm SZ}$ is a good, robust mass indicator
with low scatter in the $Y_{\rm SZ}-M$ relation, regardless of the
dynamical state of the cluster.  In addition, the $Y_{\rm SZ}-M$
relation exhibits a simple, nearly self-similar evolution with redshift
\cite{dasilva_etal04,nagai06}. The physical reason for the robustness of
the SZ flux is straightforward: $Y_{\rm SZ}$ is directly related to the
total thermal energy of the ICM and thus to the depth of the cluster
potential well. 

Here we show that a similar robust, low-scatter mass indicator can be
constructed using X-ray observables. The indicator, which is simply the
product of the X-ray derived gas mass and average temperature,
{\boldmath $Y_{\rm x}=M_{\rm g}\,\Tx$}, correlates strongly with cluster
mass with only $\approx 5-8\%$ intrinsic scatter. The $\Yx-\Mtot$
relation is robust to mergers, in the sense that even for disturbed
unrelaxed systems it gives unbiased estimates of mass with the
statistical uncertainty similar to that for relaxed systems. In
addition, we show that evolution of the slope and normalization of the
$Y_{\rm X}-M$ relation is nearly self-similar.  These properties make
$Y_{\rm X}$ particularly useful for measurements of cluster mass
function using X-ray surveys.

\section{Mass Proxies}
\label{s:indicators}

Physical properties of virialized systems, such as clusters, are
expected to correlate with their total mass. For example, in the
self-similar model \cite{kaiser86} the cluster gas mass is expected to
be simply proportional to the total mass, $M_{\Delta_c}=C_{M_g} M_{\rm
  g,\Delta_c}$, where masses are determined within a radius enclosing a
certain overdensity $\Delta_c$ with respect to the critical density of
the universe at the epoch of observation, $\rho_{\rm crit}(z)$, and
$C_{M_g}$ is a constant independent of cluster mass and redshift. The
self-similar relation between cluster mass and temperature is $
E(z)\,M_{\Delta_c}=C_T T^{3/2}.$ Here the function $E(z)\equiv H(z)/H_0$
for a flat cosmology with the cosmological constant assumed throughout
this study is given by $E(z)=\left(\Omega_M(1+z)^3 +
  \Omega_{\Lambda}\right)^{1/2}$. 

The SZ flux integrated within a certain radius, $Y_{\rm SZ}$, is
proportional to the total thermal energy of the ICM gas and thus to the
overall cluster potential, which makes it relatively insensitive to the
details of the ICM physics and merging. $Y_{\rm SZ}$ is proportional to
the ICM mass and gas mass-weighted mean temperature, $Y_{\rm SZ}\propto
M_{\rm g,\Delta_c} T_m$. The self-similar prediction for the $Y_{\rm
  SZ}-M$ relation is
\begin{equation}
\label{eq:yszm}
E(z)^{2/5}M_{\Delta_c}= C_{Y_{\rm SZ}}  Y_{\rm SZ}^{3/5}. 
\end{equation}
Cosmological simulations show that $Y_{\rm SZ}$ is a good, low-scatter
cluster mass proxy and that $Y_{\rm SZ}-M$ relation form and evolution
are close to the self-similar prediction
\cite{dasilva_etal04,hallman_etal06,nagai06}. Given the good qualities
of $Y_{\rm SZ}$ as a mass proxy, it is interesting whether a similar
indicator can be constructed from the X-ray observables, which could be
used in studies of the X-ray cluster abundances. The simplest X-ray
analog of $Y_{\rm SZ}$ is
\begin{equation}\label{eq:Yx:def}
\Yx = M_{\rm g,\Delta_c}\, \Tx,
\end{equation}
where $M_{\rm g,\Delta_c}$ is the gas mass derived from the X-ray
imaging data (it is measured within a radius enclosing overdensity
$\Delta_c$), and $\Tx$ is the mean X-ray spectral temperature. $\Tx$ is
measured excluding the central cluster region, which can be achieved
with moderate angular resolution X-ray telescopes ($\le15''$ FWHM). To
excise the central regions is desirable because the observed cluster
temperature profiles show a greater degree of similarity outside the
core \cite{vikhlinin_etal06}, and also because this makes the spectral
temperature closer to the gas mass averaged $T_m$.

%


\section{Mock \emph{Chandra} Images and Analyses of Simulated Clusters }
\label{sec:mock}

To test the quality of $T_{\rm X}$, $M_g$, and $Y_X$ as the total mass
proxies, we take the following approach. We use the output of the
high-resolution cosmological simulations of clusters in a wide mass
range to predict their X-ray emission maps. These maps are convolved
with the response of the \emph{Chandra} telescope to create realistic
mock ``observations'' of these clusters. The mock data are used as an
input to the actual X-ray data analysis pipeline and measure the proxies
($T_{\rm X}$, $M_g$, and $Y_X$) as they would be derived by observers. 

The simulations and analysis procedure is fully described in
Kravtsov~et~al.~\cite{2006astro.ph..3205K} The cosmological simulations
accurately follow the cluster growth from inital conditions using
Adaptive Refinement Tree (ART) $N$-body$+$gasdynamics code
\cite{kravtsov_etal02}, a Eulerian code that reaches the high dynamic
range required to resolve cores of mass halos. The peak formal
resolution achieved in these simulations, $\approx 3.66\,h^{-1}$~kpc, is
sufficient to resolve halos of individual galaxies and to follow not
only dissipationless dynamics of dark matter and gasdynamics of the ICM,
but also star formation, metal enrichment and thermal feedback due to
the supernovae type II and type Ia, self-consistent advection of metals,
metallicity dependent radiative cooling and UV heating due to
cosmological ionizing background.

\begin{figure*}
\centerline{
  \framebox{\includegraphics[width=0.315\linewidth]{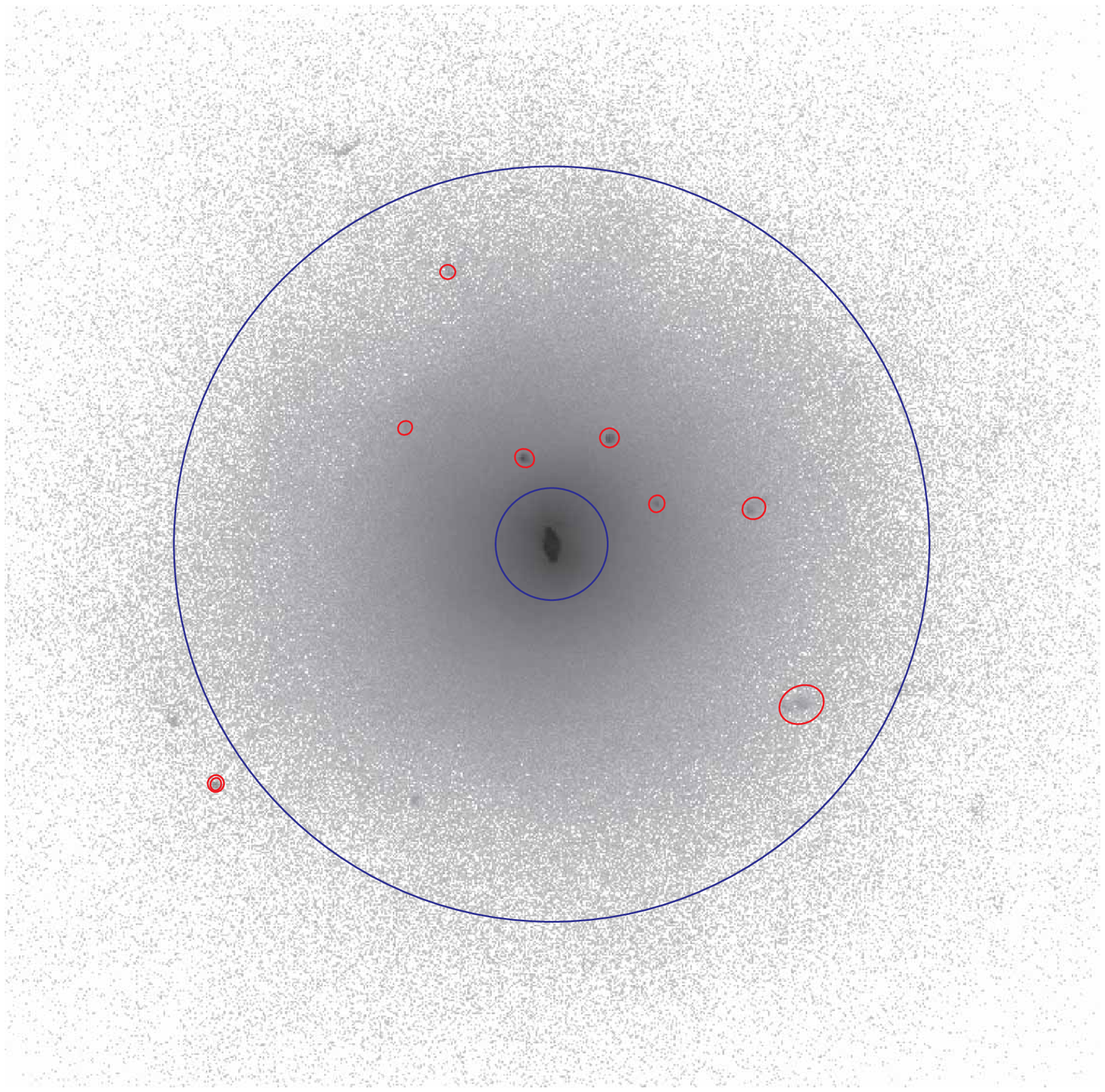}}
  \framebox{\includegraphics[width=0.315\linewidth]{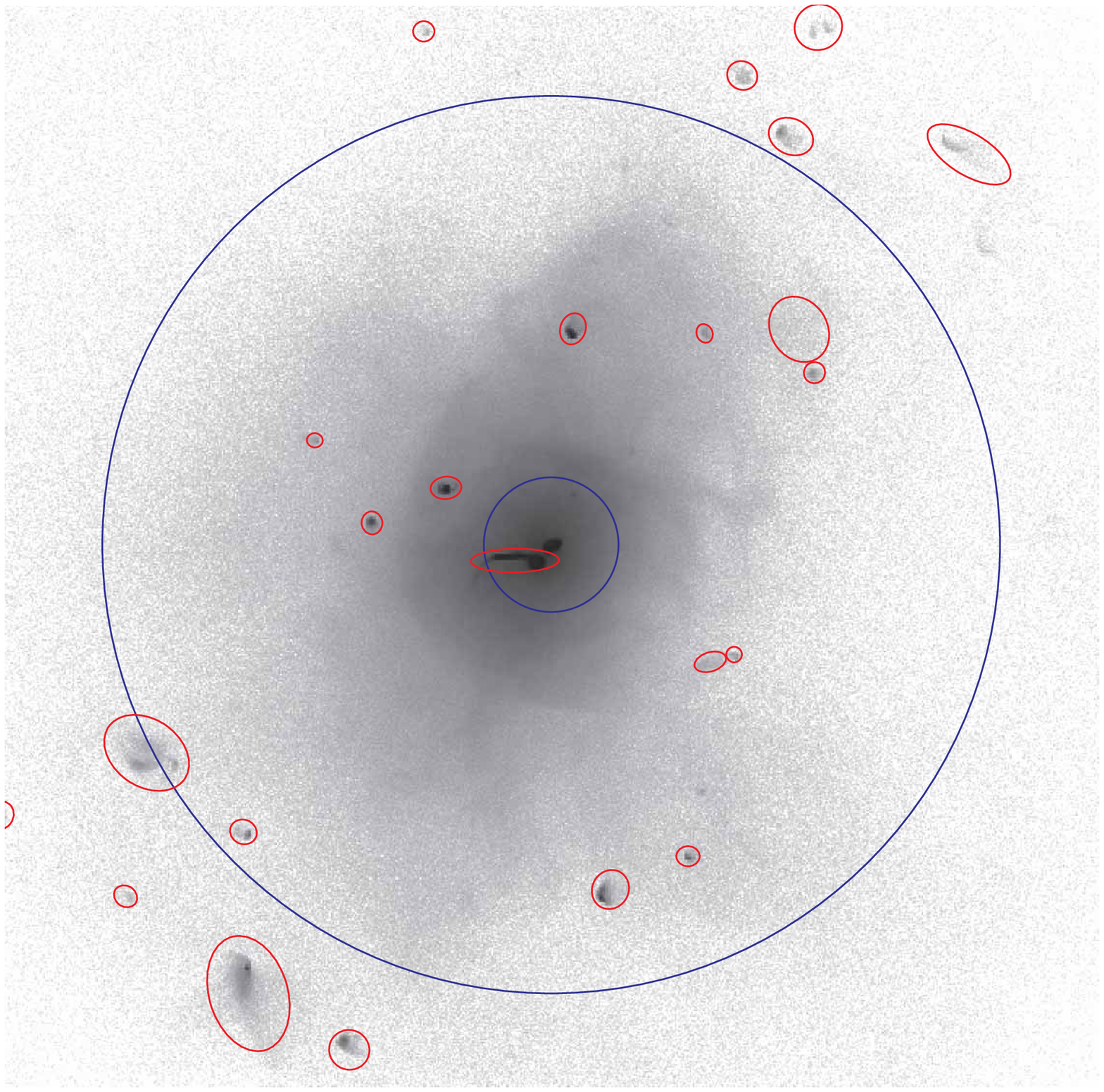}}
  \framebox{\includegraphics[width=0.315\linewidth]{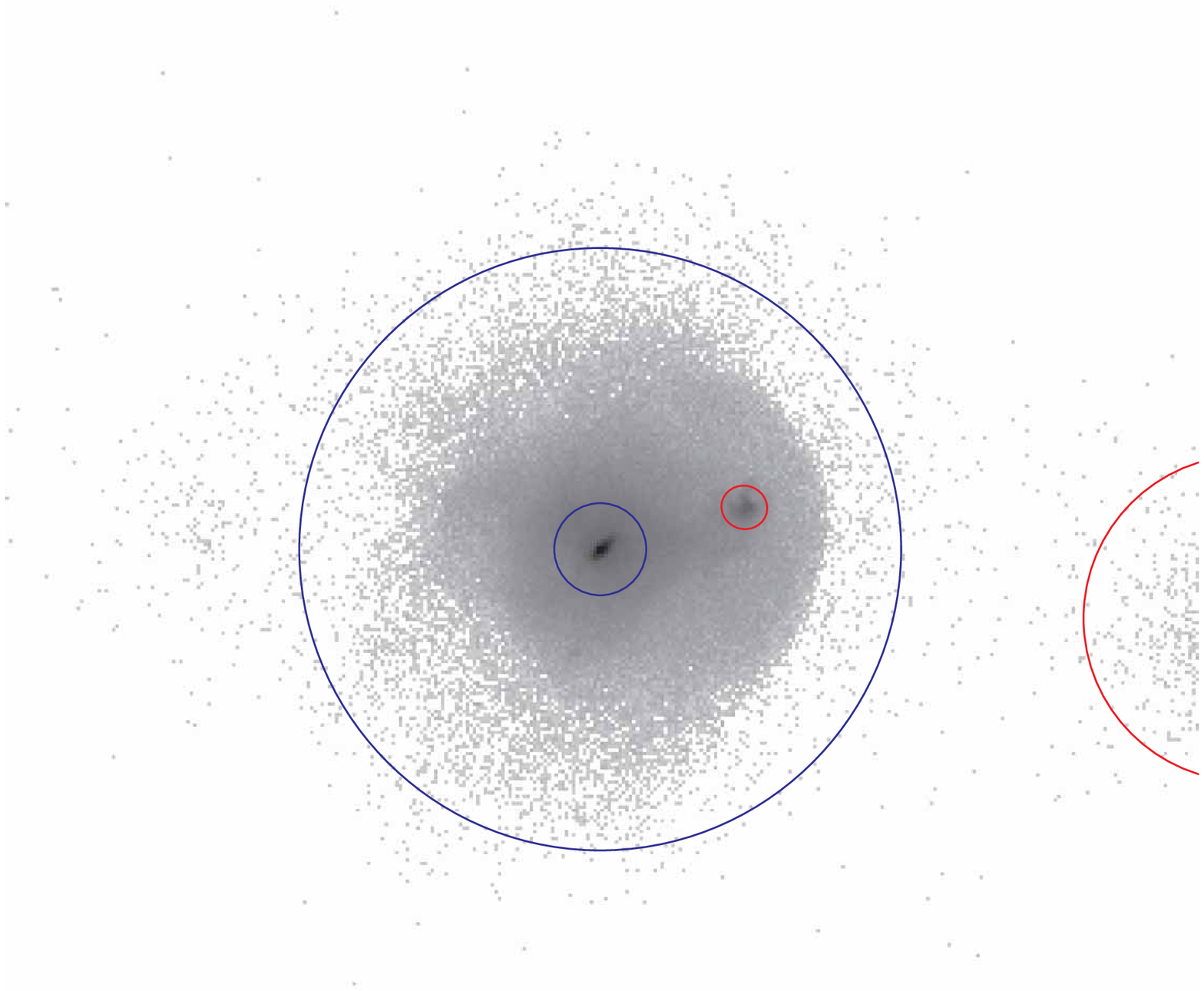}}
}
\caption{Examples of mock \emph{Chandra} images of our simulated
  clusters. Small ellipses show substructures detected by our automated
  software on the mock 100~ksec images.  Large circles show the radii
  $r=r_{500}$ and $0.15\,r_{500}$.} 
\label{fig:images}
\end{figure*}

Generation and analysis of the mock X-ray data reproduces all the
effects associated with the \emph{Chandra} response and reconstruction
of the 3D ICM properties from the observed projected X-ray spectra. The
only exception is that we ignore complications present in reduction of
the real \emph{Chandra} data related to background subtraction and
spatial variations of the effective area (i.e., we assume that accurate
corrections for these effects can be applied to the real data and any
associated uncertainties are included in the reported measurement
errors). 

Our simulations include clusters in a wide range of mass and dynamical
state. We use simulation outputs at redshifts $z=0$ and $z=0.6$ to test
the evolution in the mass~vs.~proxy relations. Examples of the mock
\emph{Chandra} images are shown in Fig.\ref{fig:images}.

\section{Comparison of Mass Indicators}
\label{s:compind}

\begin{figure*}[t]
\begin{minipage}[t]{0.48\linewidth}
\includegraphics[width=0.99\linewidth,bb=18 177 547 660]{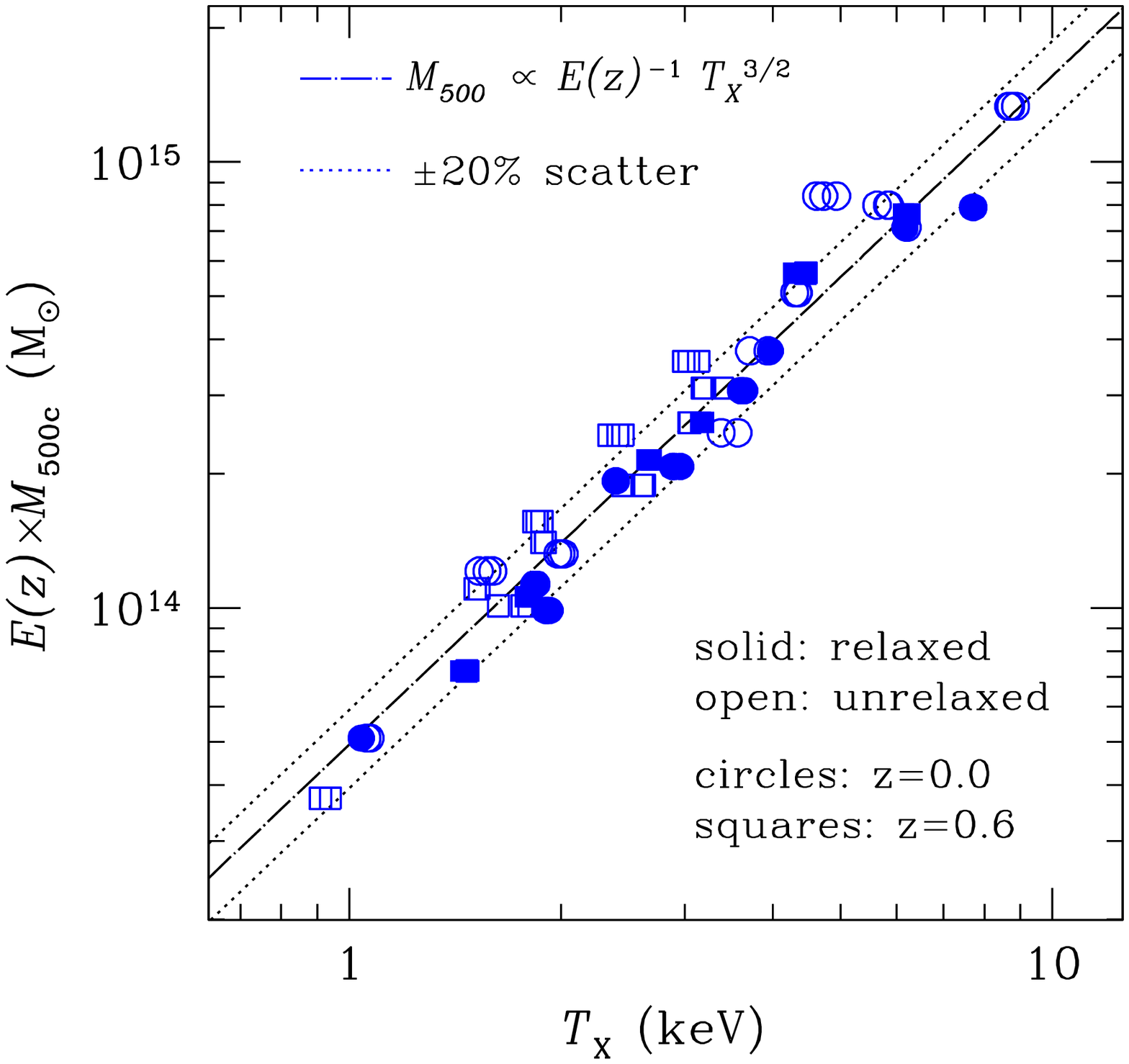}
\vspace*{-1.5\baselineskip}
\caption{Relation between the X-ray spectral temperature, $\Tx$, and
  total mass, $M_{500}$. Separate symbols indicate relaxed and unrelaxed
  clusters, and also $z=0$ and $z=0.6$ samples. The dashed line shows
  the power law relation with the self-similar slope fit to the entire
  sample, and the dotted lines indicate $20\%$ scatter. }
\label{fig:txm}
\end{minipage}
\hfill
\begin{minipage}[t]{0.48\linewidth}
\includegraphics[width=0.99\linewidth,bb=18 177 547 660]{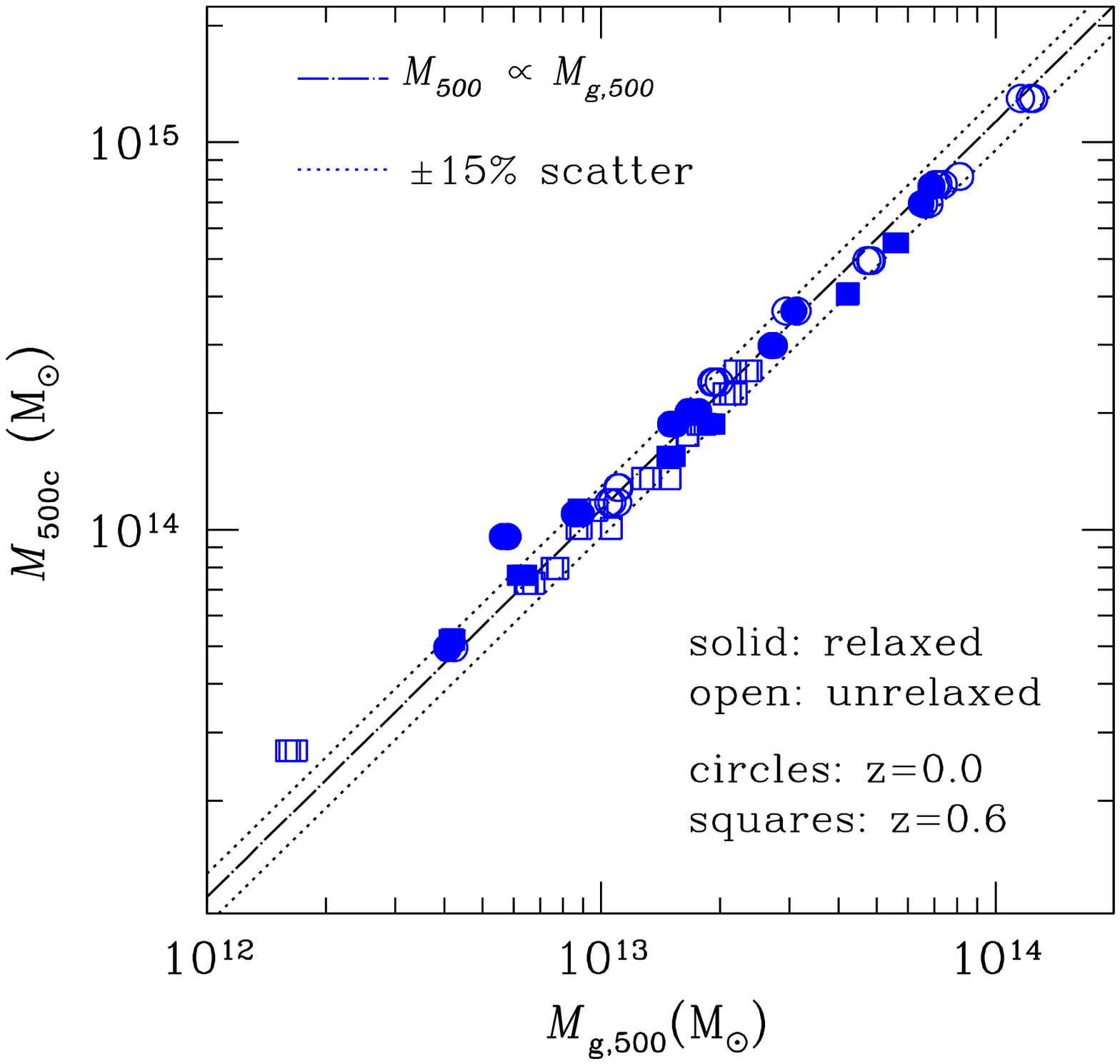}
\vspace*{-1.5\baselineskip}
\caption{Correlation between gas mass and total mass of the
clusters. Both masses are measured within $r_{500}$. The meaning of
the symbols and lines is the same as in Fig.~\ref{fig:txm}. The dotted
lines indicate $15\%$ scatter.  }
\label{fig:mgm}
\end{minipage}
\end{figure*}

Figure~\ref{fig:txm} shows that the slope and evolution of the
$\M500-\Tx$ relation\footnote{Hereafter, $\M500$ is the total mass
  within the sphere corresponding to the mean overdensity of 500
  relative to the critical density at the cluster redshift.} are quite
close to the self-similar model. There is a $\sim 20\%$ scatter in
$\M500$ around the mean relation and much of the scatter is due to
unrelaxed clusters. Note also that the normalizations of the $\M500-\Tx$
relation for relaxed and unrelaxed systems are somewhat different:
unrelaxed clusters have lower temperatures for a given mass.  This may
seem counter-intuitive at first, given that one can expect that shocks
can boost the ICM temperature during mergers.  However, in practice the
effect of shocks is relatively small~\cite{ohara_etal06}.  The main
source of the bias is that during advanced mergers the mass of the
system already increased but only a fraction of the kinetic energy of
merging systems is converted into the thermal energy of the
ICM~\cite{mathiesen_evrard01}. 

The $\M500-\Mg$ relation (Fig.~\ref{fig:mgm}) has a somewhat smaller
scatter ($\approx 10-12\%$) around the best fit power law relation than
the $\M500-\Tx$, but its slope is significantly different from the
self-similar prediction --- we find $\M500\propto \Mg^{0.88\div 0.92}$
compared to the expected $\M500\propto \Mg$. This is due to the trend of
gas fraction with cluster mass, $f_{\rm gas}\equiv \Mg/\M500\propto
\M500^{0.1\div 0.2}$ present for both the simulated clusters in our
sample~\cite{kravtsov_etal05} and for the observed
clusters~\cite{vikhlinin_etal06}.  The normalization of the $\M500-\Mg$
relation evolves only weakly between $z=0.6$ and $z=0$ (yet, the
evolution is statistically significant and it reflects slow evolution of
the gas fraction with time~\cite{kravtsov_etal05}).

\begin{figure*}
\begin{minipage}[t]{0.48\linewidth}
\includegraphics[width=0.99\linewidth,bb=18 177 547 660]{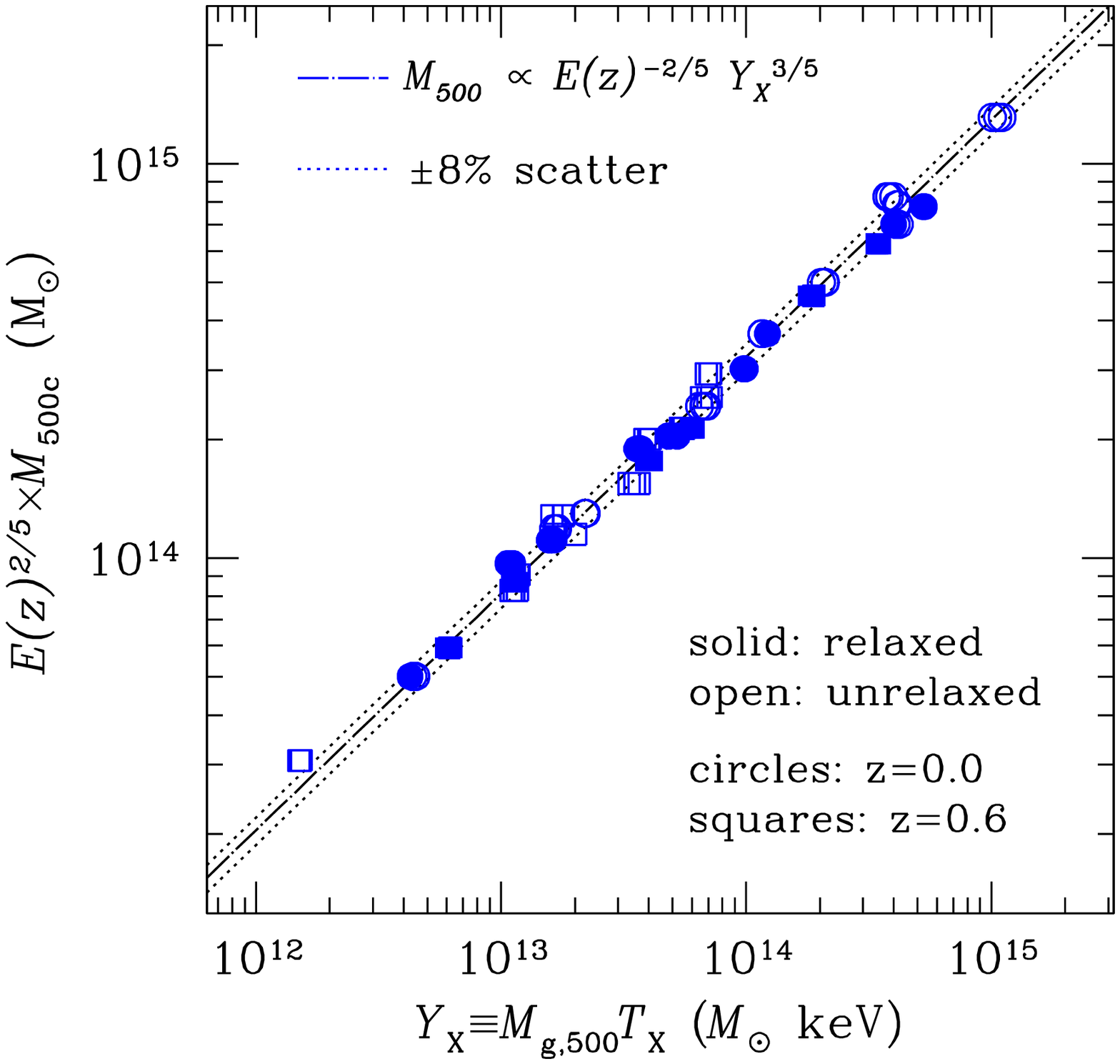}
\vspace*{-1.5\baselineskip}
\caption{$\Yx-M_{500}$ correlation. The meaning
of the symbols and lines is the same as in Fig.~\ref{fig:txm}. The 
dotted lines indicate $8\%$ scatter. } 
\label{fig:yxm}
\end{minipage}
\hfill
\begin{minipage}[t]{0.48\linewidth}
\includegraphics[width=0.99\linewidth,bb=18 177 547 660]{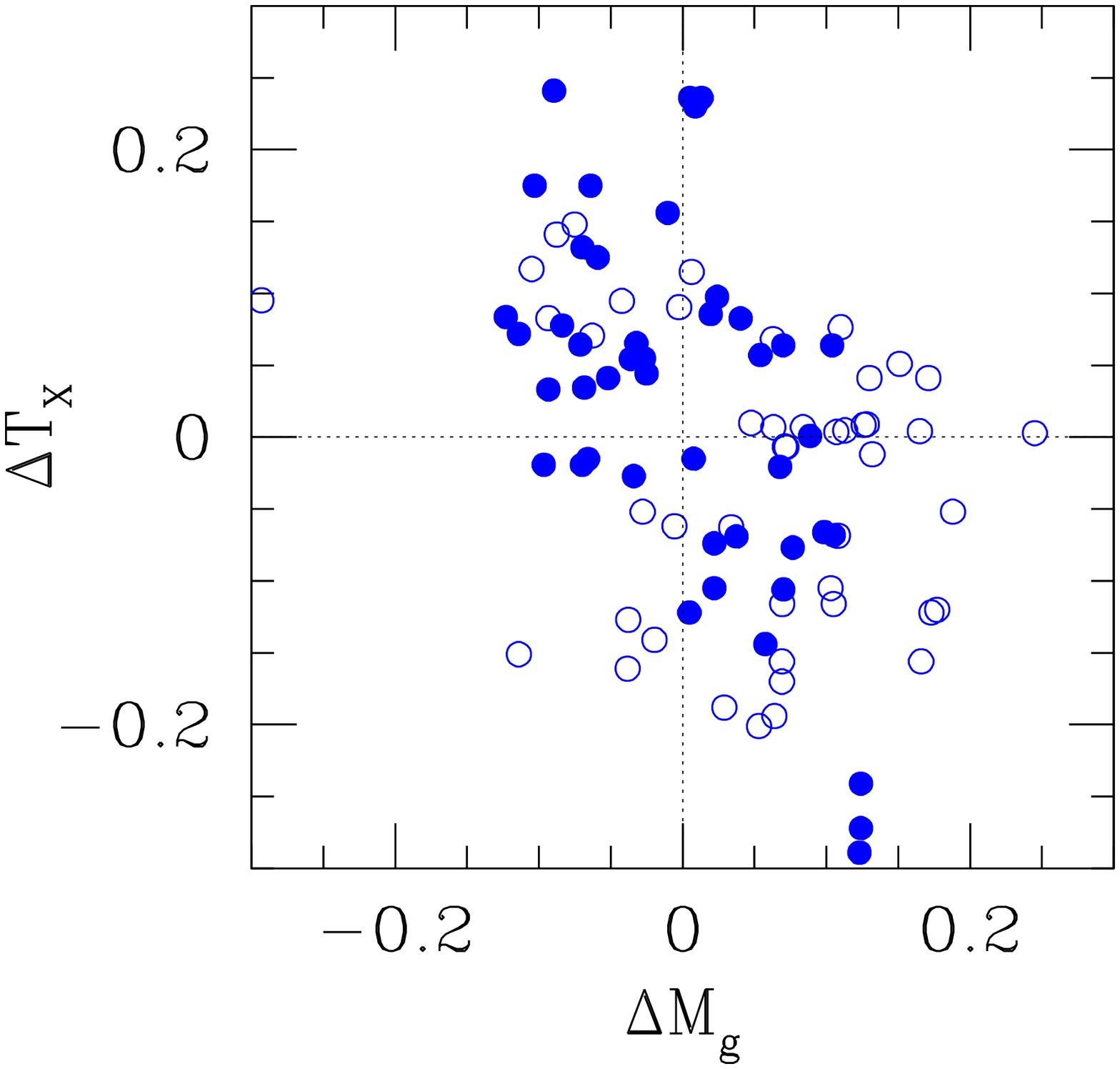}
\vspace*{-1.5\baselineskip}
\caption{Fractional deviations in $T_X$ and $M_g$ for fixed $\Mtot$ from
  their self-similar relations. Solid and open circles show clusters at
  $z=0$ and $z=0.6$, respectively.  The deviations for $M_g$ and $T_X$
  are generally anti-correlated.  A similar anti-correlation exists in
  the trend with redshift.} 
\label{fig:res}
\end{minipage}
\end{figure*}

The $\M500-\Yx$ relation (Fig.~\ref{fig:yxm}) has the smallest scatter
of only $\approx 5-7\%$. Note that this value of scatter includes
clusters at both low and high-redshifts and both relaxed and unrelaxed
systems.  In fact, the scatter in $\M500-\Yx$ for relaxed and unrelaxed
systems is indistinguishable within the errors.  Note also that the
figures include points corresponding to the three projections of each
cluster. Figure~\ref{fig:yxm} shows that the dispersion in the projected
values of $\Yx$ for each given cluster is very small, which means that
$\Yx$ is not very sensitive to the asphericity of clusters.  Remarkably,
the scatter of the $\M500-\Yx$ relation, which involves direct X-ray
observables, is as small as that in the $\M500-Y_{\rm SZ}$ relation
($\approx 7\%$ for our sample). 

The comparison of the mass proxies, clearly shows that $\Yx$, the
product of gas mass and X-ray spectral temperature, is more robust and
self-similar mass indicator than either of these X-ray observables.  Why
is the product better than its parts? The answer is obvious from
Figure~\ref{fig:res} where we show the residuals of temperature and gas
mass from their respective relations with total mass. 
Figure~\ref{fig:res} shows that the clusters with temperatures lower
than the mean temperature for a given total mass tend to have gas mass
higher than the mean, and vice versa. Note also that there is some
redshift evolution between $z=0$ and $z=0.6$ --- more clusters have
negative deviations of temperature and positive deviations of measured
gas mass at $z=0.6$ compared to $z=0$.  This redshift evolution is thus
in the opposite direction for the gas mass and temperature deviation. 
The measured $\Mg$ systematically increases at higher $z$ for a fixed
total mass because high-$z$ clusters are less relaxed on average. For
unrelaxed clusters, the ICM density distribution is non-uniform which
results in overestimation of $\Mg$ from the X-ray data
\cite{mathiesen_etal99}.  Some of the decrease of $\Mg$ at lower $z$ may
be due to continuing cooling of the ICM which decreases the mass of hot,
X-ray emitting gas.  The anti-correlation of residuals and opposite
evolution with redshift for gas mass and temperature is the reason why
the behavior of their product, on average, has smaller scatter and is
closer to the self-similar expectation in both the slope and evolution. 

\section{Discussion and Conclusions}
\label{s:discussion}

We presented comparison of several X-ray proxies for the cluster mass
--- the spectral temperature $T_x$ and gas mass $M_g$ \emph{derived from
  the X-ray data} within $r_{500}$, and the new proxy, $\Yx$, defined as
a simple product of $\Tx$ and $\Mg$.  Analogously to the integrated
Sunyaev-Zel'dovich flux, $\Yx$ is related to the total thermal energy of
the ICM. To test these mass proxies, we use mock \emph{Chandra}
``observations'' of a sample of clusters simulated in the concordance
$\Lambda$CDM cosmology. 

The main result of this study is that $\Yx$ is a robust mass indicator
with remarkably low scatter of only $\approx 5-7\%$ in $\M500$ for fixed
$\Yx$, regardless of whether the clusters are relaxed or not. In
addition, the redshift evolution of the $\Yx-M_{500}$ relation is close
to the self-similar prediction given by equation~\ref{eq:yszm}, which
makes this indicator a very attractive observable for studies of cluster
mass function with the X-ray selected samples. 

The $\Tx-\M500$ relation has the largest scatter ($\approx 20\%$), most
of which is due to unrelaxed clusters. The unrelaxed clusters have
temperatures biased low for a given mass because a certain fraction of
the kinetic energy of merging systems is still in the form of bulk
motions of the ICM.  The $\Mg-\M500$ relation shows an intermediate
level of scatter, $\approx 10-12\%$. This relation does not appear to be
sensitive to mergers. It does, however, exhibit significant deviations
from self-similarity in its slope, which is due to the dependence of gas
fraction within $r_{500}$ on the cluster mass \cite{kravtsov_etal05} (a
similar dependence exists for the observed clusters
\cite{vikhlinin_etal06}). 

Generally, all the observable--mass relations we tested demonstrate a
remarkable degree of regularity of galaxy clusters as a population. $\Tx$,
$\Mg$, and $\Yx$ all exhibit correlations with $\M500$ which are close to
the expectation of the self-similar model, both in their slope and
evolution with time, within the uncertainties provided by our
sample. The only exception is the slope of the $\Mg-\M500$ relation. 

\begin{figure*}[t]
\begin{minipage}[t]{0.48\linewidth}
\includegraphics[width=0.97\linewidth,bb=18 177 547 670]{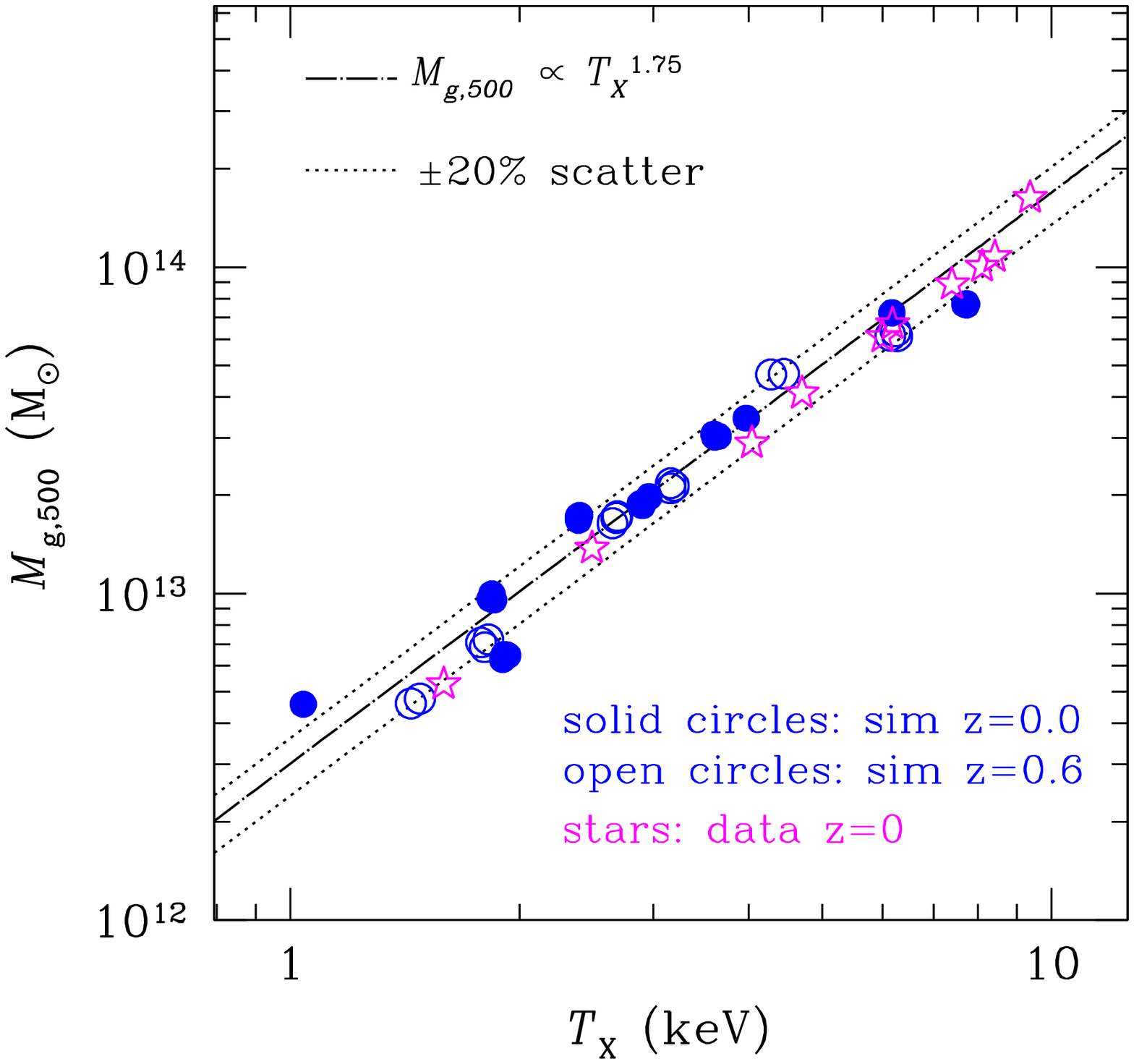}
\caption{Relation between X-ray spectral temperature and gas mass for
  the relaxed subsample of simulated clusters (circles) and for a sample
  of relaxed \emph{Chandra} clusters (\emph{stars}).  Both gas mass and
  temperature are the quanities derived from analysis of real and mock
  X-ray data. The error bars in the \emph{Chandra} measurements are
  comparable to the symbol size and are not shown for clarity. The
  \emph{dashed line} shows the best fit power law relation with the
  slope $1.75$.} 
\label{fig:tmg}
\end{minipage}
\hfill
\begin{minipage}[t]{0.48\linewidth}
\includegraphics[width=0.97\linewidth,bb=18 177 547 670]{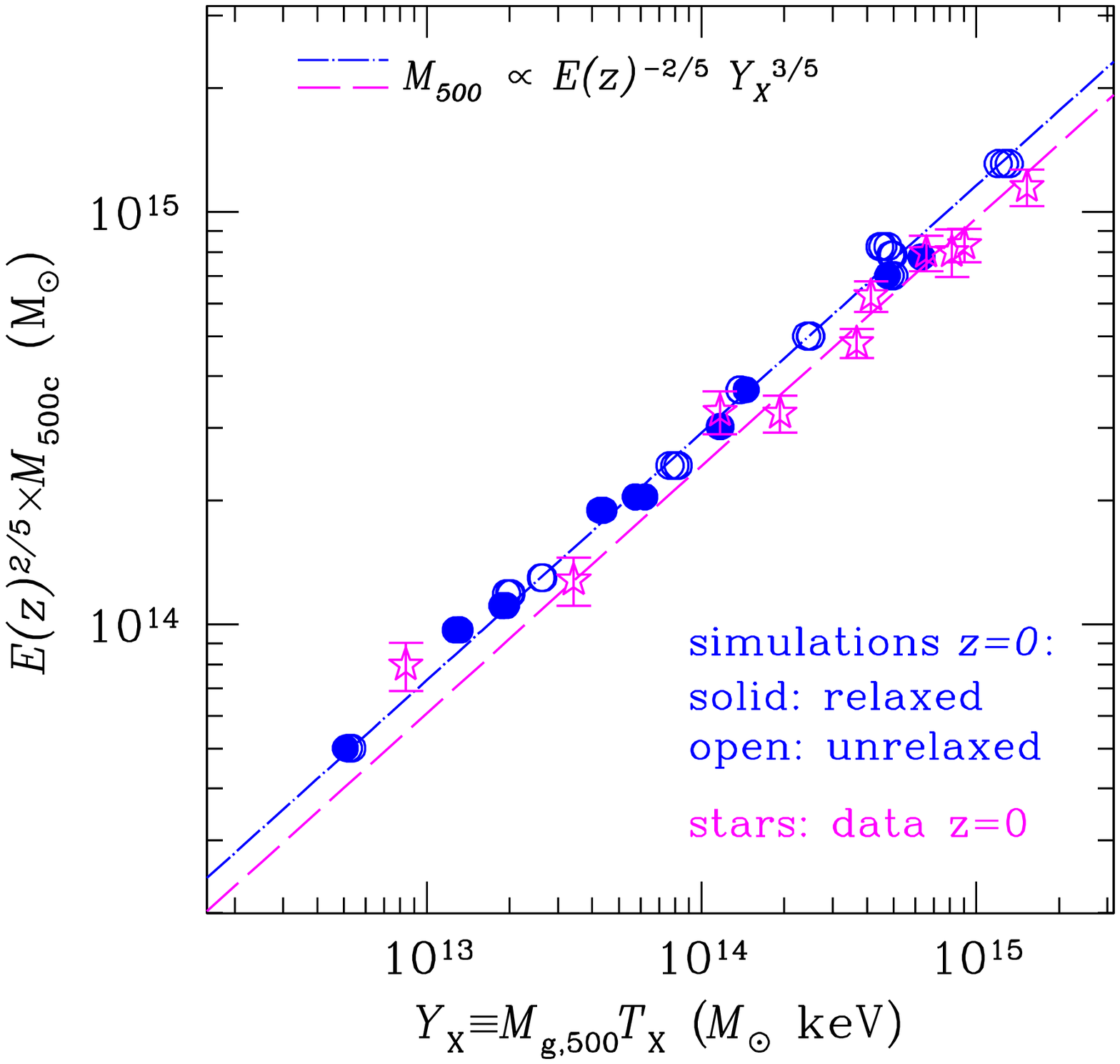}
\caption{$\Yx-\M500$ relation for the $z=0$ sample of the simulated
  clusters (\emph{circles}) and for a sample of relaxed \emph{Chandra}
  clusters.  The gas masses for the simulated clusters are appropriately
  rescaled (see caption to Fig.\ref{fig:tmg}). The {\it dot-dashed} line
  shows the best fit power law relation for the simulated clusters with
  the slope fixed to the self-similar value of $3/5$. The {\it dashed}
  line shows the same best fit power law, but with the normalization
  scaled down by $15\%$. }
\label{fig:yxmc}
\end{minipage}
\end{figure*}

Given that our analysis relies on cosmological simulations, it is
reasonable to ask whether the simulated clusters are realistic. 
Although simulations certainly do not reproduce all of the observed
properties of clusters, especially in their core regions, the ICM
properties outside the core in simulations and observations agree quite
well. We illustrate this in Fig.~\ref{fig:tmg}, which shows that the
$\Mg-\Tx$ relations for simulated and observed clusters
\cite{vikhlinin_etal06}. Clearly, both simulated and observed clusters
exhibit tight correlations between $\Mg$ and $\Tx$ which agree
remarkably in their slope ($M_g \propto T^{1.75}$) and normalization. 
The normalizations derived from simulated and real clusters agree to
$\approx 10\%$, while slopes are indistiguishable and both deviate
significantly from the expected self-similar value of $1.5$. This is a
consequence of significant trends in the gas fraction with cluster mass,
$\Mg/M_{500}\propto M_{500}^{0.2\div 0.25}$ for both simulated
\cite{kravtsov_etal05} and observed clusters \cite{vikhlinin_etal06}. 
The deviations from the self-similar model also manifest themselves in
the absence of any noticeable evolution with redshift\footnote{Note that
  $\Mg$ in Fig.\ref{fig:tmg} is not multiplied by the $E(z)$ factor
  unlike the total mass in Fig.\ref{fig:txm} and \ref{fig:yxm}.}. 
Interestingly, the real clusters show a similarly weak evolution in the
$\Mg-\Tx$ relation \cite{vikhlinin_etal02}. Figure~\ref{fig:res} shows
that the likely explanation is that the clusters at $z=0.6$ tend to be
colder for the fixed $\Mtot$ but have higher estimated $\Mg$ then their
counterparts at $z=0$ because they are less relaxed. 

A similar level of agreement between the simulations and latest
\emph{Chandra} measurements exists also for the total mass vs.{}
temperature relation, $M_{500}-\Tx$. In fact, the normalization for our
simulated sample agrees with the observational
results\cite{vikhlinin_etal06} to $\approx 10\%$. This is a considerable
improvement over the situation of just several years ago when there was
$\approx 30-50\%$ discrepancy between observational measurements and
cosmological simulations~\cite{finoguenov_etal01,pierpaoli_etal03}. The
$M-\Tx$ normalization was revised both in simulations and observations
due to (1) inclusion of more realistic physics in cosmological
simulations (e.g., radiative cooling and star formation,
\cite{dave_etal02,muanwong_etal02}, (2) improved analyses of observed
clusters using more realistic gas density profiles
\cite{borgani_etal04,vikhlinin_etal06}, (3) more reliable measurements
of the cluster temperature profiles
\cite{markevitch_etal98,arnaud_etal05,vikhlinin_etal06}, and (4) the use
of uniform definition of $\Tx$ in observations and in simulations
analyses \cite{mazzotta_etal04,rasia_etal05,vikhlinin06}. The remaining
systematic $10\%$ difference observed at present is likely caused by
non-thermal pressure support from bulk gas motions
\cite{faltenbacher_etal05,rasia_etal06,lau_etal06}, which is unaccounted
for by the X-ray hydrostatic mass estimates. In Figure~\ref{fig:yxmc} we
compare the $\Yx-\M500$ relation for the simulated clusters and for the
\emph{Chandra} sample\cite{vikhlinin_etal06}. The observed clusters show
a tight correlation with the slope close to the self-similar value. 
There is $\approx 15\%$ difference in normalization, likely explained
also by neglecting the turbulent pressure support in the \emph{Chandra}
hydrostatic mass estimates. The excellent agreement of simulations and
observations in terms of the relation between the two X-ray observables
used to compute $\Yx$ ($\Mg-\Tx$) and a relatively good agreement in the
$\Tx-\M500$ and $\Yx-\M500$ relations, gives us confidence that the
results presented here are sufficiently realistic. 

Our results show that $\Yx$ is clearly most robust and most self-similar
X-ray cluster mass indicator. The biases existing in mass estimates
based on $\Mg$ and $\Tx$ anti-correlate both for a given redshift and in
terms of evolutionary trends (see Figure~\ref{fig:res}). This explains
why their product, $\Yx$, is a better mass indicator than $\Tx$ and
$\Mg$ individually. The quality of $\Yx$ compares well to that for the
actual three-dimension integral of the ICM thermal energy (proportional
to $Y_{\rm SZ}$) in terms of its low scatter and self-similarity.  $\Yx$
may prove to be an even better mass proxy than $Y_{\rm SZ}$, given that
we use ideal 3D measurement of the latter while reproducing the actual
data analysis for the former.  Note also that $Y_{\rm SZ}$ is more
sensitive to the outskirts of clusters, because it involves gas
mass-weighted temperature (as opposed to the spectral temperature more
sensitive to the inner regions), and thus should be more prone to
projection effects. 

Note that $\Yx$ is also an attractive mass proxy from the data analysis
point of view. First, it reduces observational statistical noise by
combining the two independently measured quantities, $\Mg$ and $\Tx$,
into a single quantity. For example, a $10\%$ measurement uncertainty in
$\Tx$ translates into a $\sim 15\%$ mass uncertainty through the $M-\Tx$
relation and only $6\%$ uncertainty through the $\Yx-M$ relation. $\Yx$
is also less sensitive to any errors in the absolute calibration of the
X-ray telescope because the biases in the derived $T_X$ and $M_g$ tend
to anticorrelate.

The robustness and low scatter make $\Yx$ an excellent mass indicator
for observational measurements of cluster mass function at both $z=0$
and higher redshifts. The necessary data --- an X-ray brightness profile
and a wide-beam spectrum excluding the core --- are easily obtained with
sufficiently deep observations with \emph{Chandra}, \emph{XMM-Newton},
and \emph{Suzaku} (for low-redshift clusters). The small scatter and
simple, nearly self-similar evolution of the $\Yx-M$ relation hold
promise for the self-calibration strategies for future large X-ray
cluster surveys. 

\bigskip

This project was supported by the NSF under grants No.{} AST-0206216 and
AST-0239759, and by NASA through grants NAG5-13274 \& NAG5-9217 and
contract NAS8-39073. D.N. is supported by a Sherman Fairchild
postdoctoral fellowship at Caltech. Last but not least, we would like to
thank the organizers of the XLIst \emph{Rencontres du Moriond}. 

\bibliography{ms}

\begin{thebibliography}{10}

\bibitem{2006astro.ph..3205K}
A.~V. {Kravtsov}, A.~{Vikhlinin}, and D.~{Nagai}.
\newblock {\em {\apj}, in press (astro-ph/0603205)}.

\bibitem{voit05}
G.~M. {Voit}.
\newblock {\em Reviews of Modern Physics}, 77:207, 2005.

\bibitem{david_etal93}
L.~{David}, A.~{Slyz}, C.~{Jones}, W.~{Forman}, S.. {Vrtilek}, and K.~{Arnaud}.
\newblock {\em \apj}, 412:479, 1993.

\bibitem{stanek_etal06}
R.~{Stanek}, A.~E. {Evrard}, H.~{B{\"o}hringer}, P.~{Schuecker}, and B.~{Nord}.
\newblock {\em {\apj} submitted (astro-ph/0602324)}, 2006.

\bibitem{allen_etal03}
S.~W. {Allen}, R.~W. {Schmidt}, A.~C. {Fabian}, and H.~{Ebeling}.
\newblock {\em \mnras}, 342:287, 2003.

\bibitem{henry_arnaud91}
J.~P. {Henry} and K.~A. {Arnaud}.
\newblock {\em \apj}, 372:410, 1991.

\bibitem{oukbir_blanchard92}
J.~{Oukbir} and A.~{Blanchard}.
\newblock {\em \aap}, 262:L21, 1992.

\bibitem{markevitch98}
M.~{Markevitch}.
\newblock {\em \apj}, 504:27, 1998.

\bibitem{ikebe_etal02}
Y.~{Ikebe}, T.~{Reiprich}, H.~{B{\"o}hringer}, Y.~{Tanaka}, and T.~{Kitayama}.
\newblock {\em \aap}, 383:773, 2002.

\bibitem{vikhlinin_etal06}
A.~{Vikhlinin}, A.V. {Kravtsov}, W.~{Forman}, C.~{Jones}, M.~{Markevitch}, S.S.
  {Murray}, and L.~{Van~Speybroeck}.
\newblock {\em {\apj}}, 640:691, 2006.

\bibitem{ohara_etal06}
T.~B. {O'Hara}, J.~J. {Mohr}, J.~J. {Bialek}, and A.~E. {Evrard}.
\newblock {\em \apj}, 639:64, 2006.

\bibitem{vikhlinin_etal03}
A.~{Vikhlinin et al.}
\newblock {\em \apj}, 590:15, 2003.

\bibitem{voevodkin_vikhlinin04}
A.~{Voevodkin} and A.~{Vikhlinin}.
\newblock {\em \apj}, 601:610, 2004.

\bibitem{levine_etal02}
E.~S. {Levine}, A.~E. {Schulz}, and M.~{White}.
\newblock {\em \apj}, 577:569, 2002.

\bibitem{hu03}
W.~{Hu}.
\newblock {\em \prd}, 67(8):081304, 2003.

\bibitem{majumdar_mohr03}
S.~{Majumdar} and J.~J. {Mohr}.
\newblock {\em \apj}, 585:603, 2003.

\bibitem{lima_hu05}
M.~{Lima} and W.~{Hu}.
\newblock {\em \prd}, 72(4):043006, 2005.

\bibitem{wang_etal04}
S.~{Wang}, J.~{Khoury}, Z.~{Haiman}, and M.~{May}.
\newblock {\em \prd}, 70(12):123008, 2004.

\bibitem{motl_etal05}
P.~M. {Motl}, E.~J. {Hallman}, J.~O. {Burns}, and M.~L. {Norman}.
\newblock {\em \apjl}, 623:L63, 2005.

\bibitem{nagai06}
D.~{Nagai}.
\newblock {\em {\apj} submitted (astro-ph/0512208)}, 2006.

\bibitem{dasilva_etal04}
A.~C. {da Silva}, S.~T. {Kay}, A.~R. {Liddle}, and P.~A. {Thomas}.
\newblock {\em \mnras}, 348:1401, 2004.

\bibitem{kaiser86}
N.~{Kaiser}.
\newblock {\em \mnras}, 222:323, 1986.

\bibitem{hallman_etal06}
E.~J. {Hallman}, P.~M. {Motl}, J.~O. {Burns}, and M.~L. {Norman}.
\newblock {\em {\apj} submitted (astro-ph/0509460)}, 2006.

\bibitem{kravtsov_etal02}
A.~V. {Kravtsov}, A.~{Klypin}, and Y.~{Hoffman}.
\newblock {\em \apj}, 571:563, 2002.

\bibitem{mathiesen_evrard01}
B.~F. {Mathiesen} and A.~E. {Evrard}.
\newblock {\em \apj}, 546:100, 2001.

\bibitem{kravtsov_etal05}
A.~V. {Kravtsov}, D.~{Nagai}, and A.~A. {Vikhlinin}.
\newblock {\em \apj}, 625:588, 2005.

\bibitem{mathiesen_etal99}
B.~{Mathiesen}, A.~E. {Evrard}, and J.~J. {Mohr}.
\newblock {\em \apjl}, 520:L21, 1999.

\bibitem{vikhlinin_etal02}
A.~{Vikhlinin}, L.~{VanSpeybroeck}, M.~{Markevitch}, W.~R. {Forman}, and
  L.~{Grego}.
\newblock {\em \apjl}, 578:L107, 2002.

\bibitem{finoguenov_etal01}
A.~{Finoguenov}, T.~H. {Reiprich}, and H.~{B{\"o}hringer}.
\newblock {\em \aap}, 368:749, 2001.

\bibitem{pierpaoli_etal03}
E.~{Pierpaoli}, S.~{Borgani}, D.~{Scott}, and M.~{White}.
\newblock {\em \mnras}, 342:163, 2003.

\bibitem{dave_etal02}
R.~{Dav{\' e}}, N.~{Katz}, and D.~H. {Weinberg}.
\newblock {\em \apj}, 579:23, 2002.

\bibitem{muanwong_etal02}
O.~{Muanwong}, P.~A. {Thomas}, S.~T. {Kay}, and F.~R. {Pearce}.
\newblock {\em \mnras}, 336:527, 2002.

\bibitem{borgani_etal04}
S.~{Borgani et al.}
\newblock {\em \mnras}, 348:1078--1096, 2004.

\bibitem{markevitch_etal98}
M.~{Markevitch}, W.~R. {Forman}, C.~L. {Sarazin}, and A.~{Vikhlinin}.
\newblock {\em \apj}, 503:77, 1998.

\bibitem{arnaud_etal05}
M.~{Arnaud}, E.~{Pointecouteau}, and G.~W. {Pratt}.
\newblock {\em \aap}, 441:893, 2005.

\bibitem{mazzotta_etal04}
P.~{Mazzotta}, E.~{Rasia}, L.~{Moscardini}, and G.~{Tormen}.
\newblock {\em \mnras}, 354:10, 2004.

\bibitem{rasia_etal05}
E.~{Rasia}, P.~{Mazzotta}, S.~{Borgani}, L.~{Moscardini}, K.~{Dolag},
  G.~{Tormen}, A.~{Diaferio}, and G.~{Murante}.
\newblock {\em \apjl}, 618:L1, 2005.

\bibitem{vikhlinin06}
A.~{Vikhlinin}.
\newblock {\em {\apj}}, 640:710, 2006.

\bibitem{faltenbacher_etal05}
A.~{Faltenbacher}, A.~V. {Kravtsov}, D.~{Nagai}, and S.~{Gottl{\"o}ber}.
\newblock {\em \mnras}, 358:139, 2005.

\bibitem{rasia_etal06}
E.~{Rasia et al.}
\newblock {\em astro-ph/0602434}, 2006.

\bibitem{lau_etal06}
E.~{Lau}, A.~V. {Kravtsov}, and D.~{Nagai}.
\newblock {\em {\apj} in preparation}, 2006.

\end{thebibliography}

\end{document}